\documentclass[]{lni}

\usepackage[caption = false]{subfig}

\newcounter{findingscounter}

\newcommand{\pred}{BLL$_{I,S}$}
\newcommand{\rec}{BLL$_{I,S,C}$}
\newcommand{\rand}{\textit{Random}}
\newcommand{\res}{\textit{CompSci}}
\newcommand{\sone}{\textit{Scenario 1}}
\newcommand{\stwo}{\textit{Scenario 2}}

\newcommand{\para}[1]{\vspace{1mm}\noindent\textbf{#1}}

\begin{document}
\title[A Cognitive-Inspired Hashtag Recommendation Approach]{The Impact of Time on Hashtag Reuse in Twitter: A Cognitive-Inspired Hashtag Recommendation Approach}
\subtitle{Presentation of work originally published in the Proc. of the 26th Intl. Conf. on WWW} 
\author[Elisabeth Lex \and Dominik Kowald]
{Elisabeth Lex\footnote{Graz University of Technology, ISDS, Inffeldgasse 13/V, 8010 Graz,
Austria \email{elisabeth.lex@tugraz.at}} \and
Dominik Kowald\footnote{Know-Center, Social Computing Group, Inffeldgasse 13/VI, 8010 Graz, Austria
\email{dkowald@know-center.at}}}
\year{2019}
\maketitle

\begin{abstract}
In our work~\cite{kowald2017www}, we study temporal usage patterns of Twitter hashtags, and we use the Base-Level Learning (BLL) equation from the cognitive architecture ACT-R~\cite{anderson2004integrated} to model how a person reuses her own, individual hashtags as well as hashtags from her social network. The BLL equation accounts for the time-dependent decay of item exposure in human memory. According to BLL, the usefulness of a piece of information (e.g., a hashtag) is defined by how frequently and how recently it was used in the past, following a time-dependent decay that is best modeled with a power-law distribution. We used the BLL equation in our previous work to recommend tags in social bookmarking systems \cite{Kowald2016a}. Here~\cite{kowald2017www}, we adopt the BLL equation to model temporal reuse patterns of individual (i.e., reusing own hashtags) and social hashtags (i.e., reusing hashtags, which has been previously used by a followee) and to build a cognitive-inspired hashtag recommendation algorithm. We demonstrate the efficacy of our approach in two empirical social networks crawled from Twitter, i.e., \res{} and \rand{} (for details about the datasets, see~\cite{kowald2017www}). Our results show that our approach can outperform current state-of-the-art hashtag recommendation approaches. 

\begin{figure*}[h]
   \centering
     \captionsetup[subfigure]{justification=centering}
     \subfloat[][Individual hashtag reuse\\$R^2$ = .883\\\res{} dataset]{ 
      \includegraphics[width=0.24\textwidth]{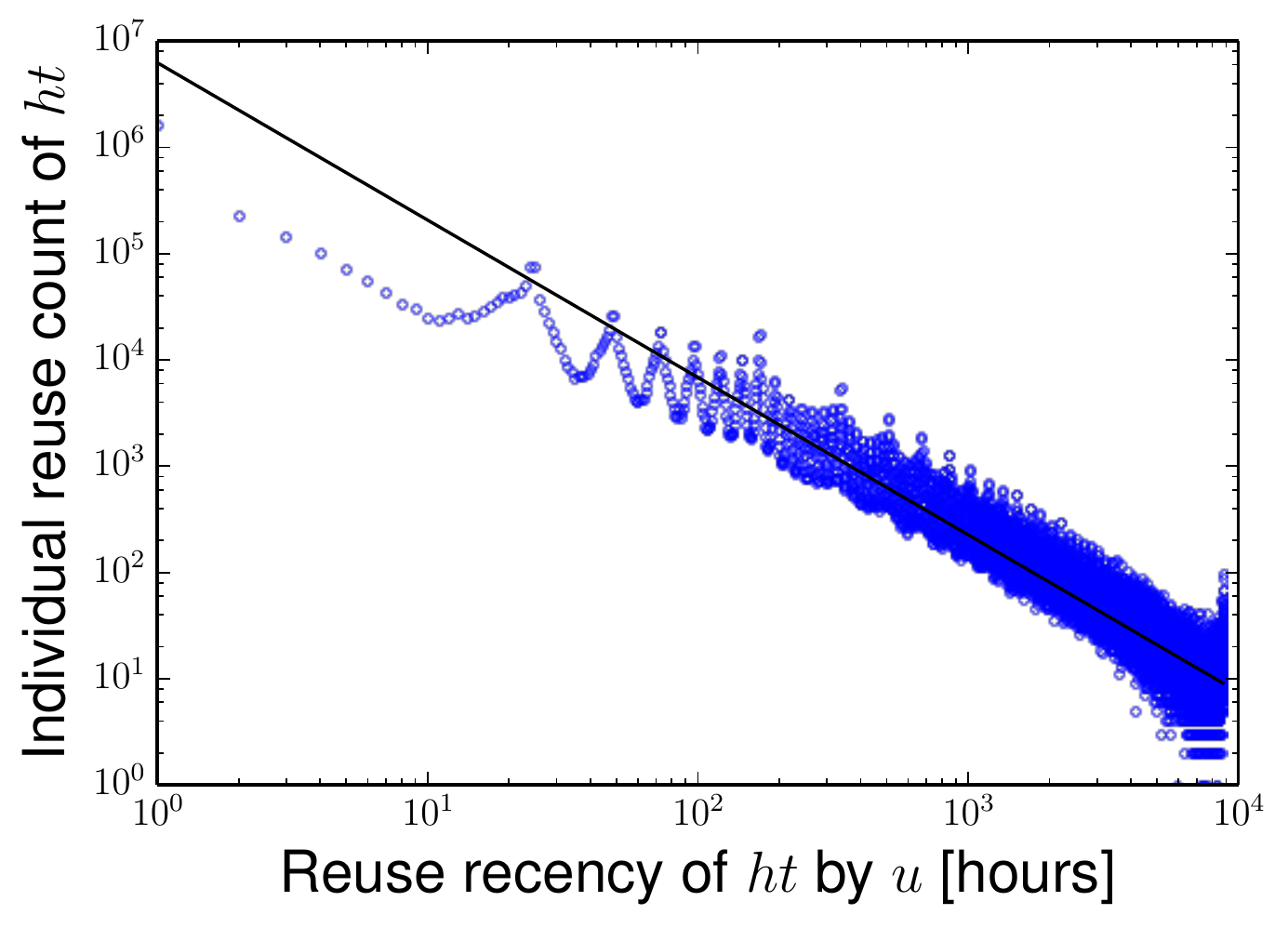} 
   }
     \subfloat[][Individual hashtag reuse\\$R^2$ = .894\\\rand{} dataset]{ 
      \includegraphics[width=0.24\textwidth]{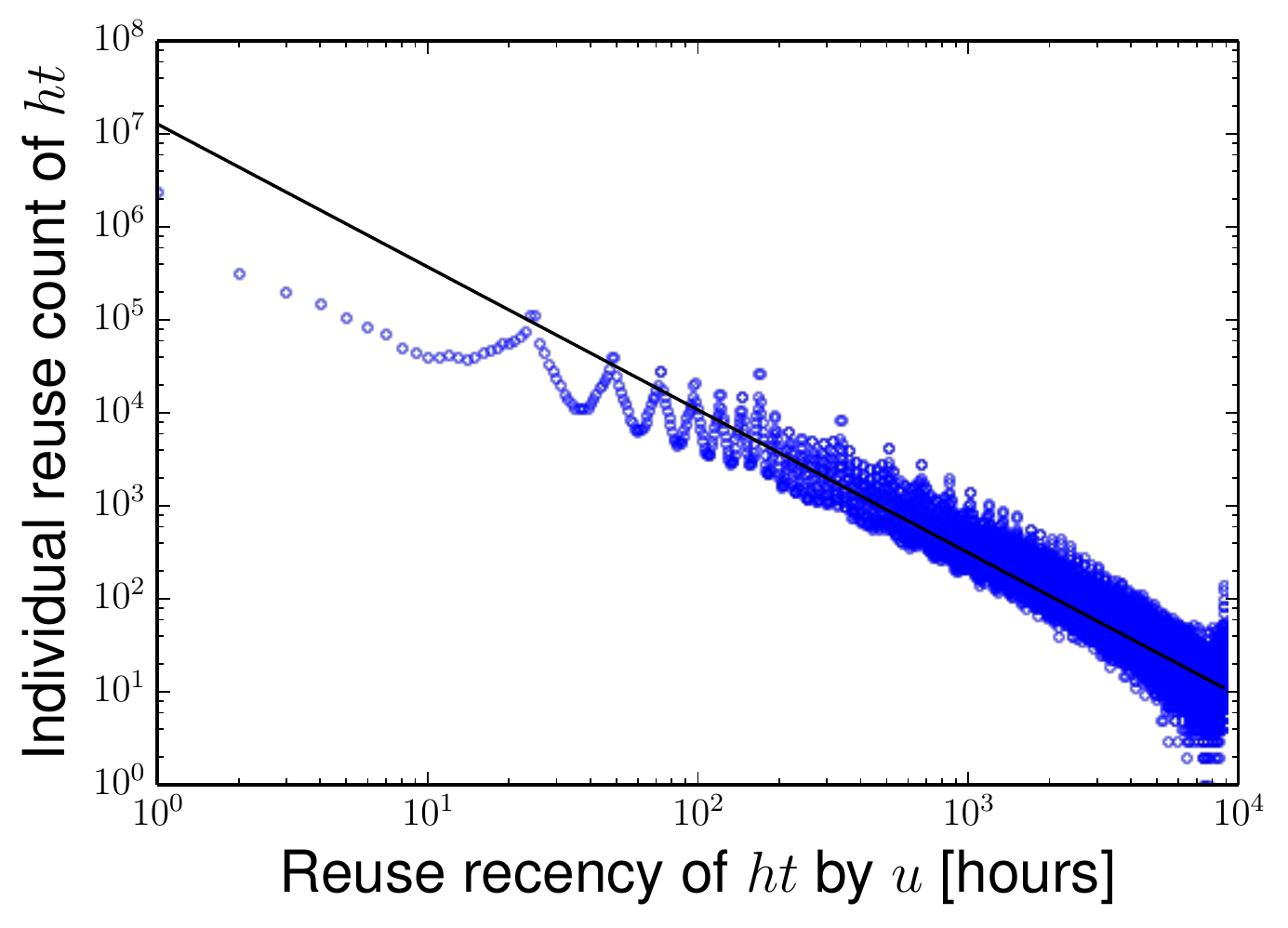} 
   }
     \subfloat[][Social hashtag reuse\\$R^2$ = .689\\\res{} dataset]{ 
      \includegraphics[width=0.24\textwidth]{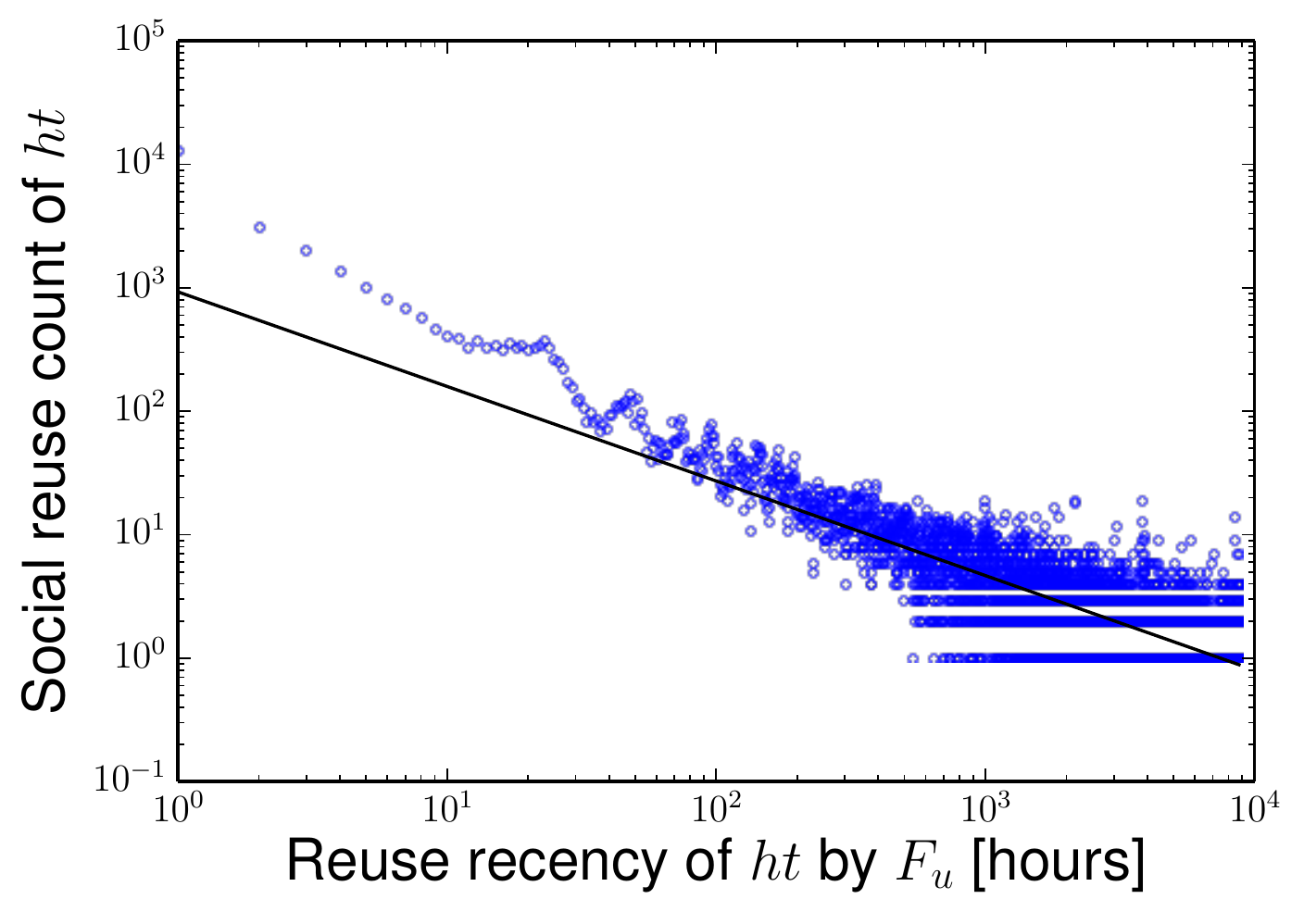} 
   }  
     \subfloat[][Social hashtag reuse\\$R^2$ = .771\\\rand{} dataset]{ 
      \includegraphics[width=0.24\textwidth]{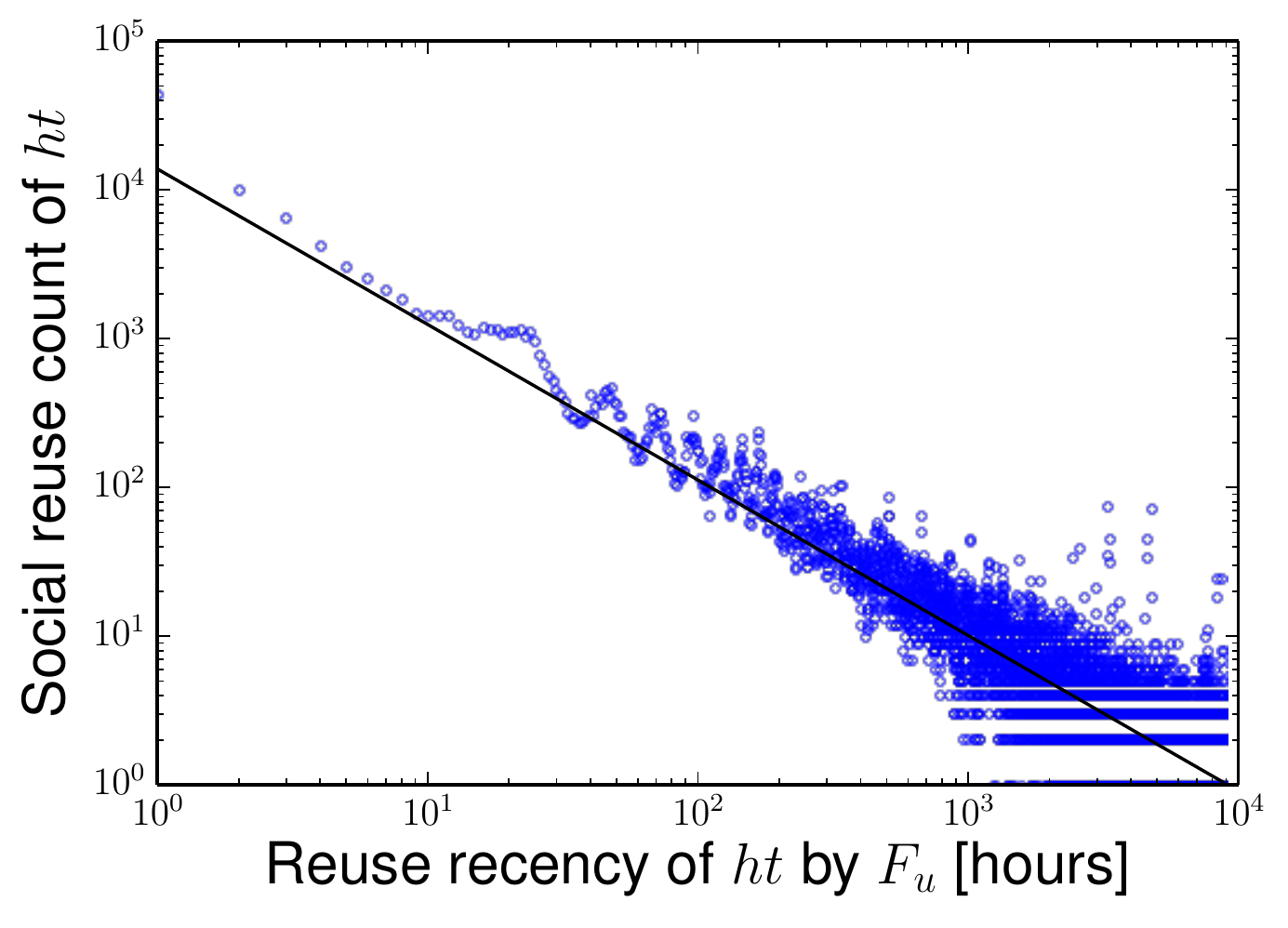} 
   }
   \caption{The effect of time on individual and social hashtag reuse (plots are in log-log scale).
\vspace{-3mm}}
     \label{fig:analysis}
\end{figure*}

\para{Temporal Effects of Hashtag Reuse.} To determine the plausibility of our approach, we study hashtag use in our two empirical Twitter datasets, i.e., \res{} and \rand{}. For each hashtag assignment, we investigate whether the hashtag has either been used by the same user before (``individual''), by some of her followees (``social''), by both (``individual/social''), by anyone else in the dataset (``network'') or by neither of them (``external''). We find that depending on the dataset, individual or social hashtag reuse can explain approximately two-third of hashtag assignments. Also, both reuse types follow a time-dependent decay that follows a power-law distribution, as shown in Figure~\ref{fig:analysis}. This motivates our idea to model both individual as well as social hashtag reuse and to recommend hashtags for new tweets with the BLL equation.

\para{Experiments and Results.} We implement the BLL equation in two variants, where the first one (i.e., \pred{}) predicts the hashtags of a user solely based on past hashtag usage, and the second one (i.e., \rec{}) combines \pred{} with a content-based tweet analysis to also incorporate the text of the currently proposed tweet of a user. We evaluate our approach using standard evaluation protocols and metrics, and we find that our approach provides significantly higher prediction accuracy and ranking estimates than current state-of-the-art hashtag recommendation algorithms in both scenarios (for more details about the baselines, refer to~\cite{kowald2017www}), as shown in Figure \ref{fig:results}.

\begin{figure*}[h]
   \centering
     \captionsetup[subfigure]{justification=centering}
     \subfloat[][\sone{}: Hashtag rec. w/o current tweet\\\res{} dataset]{ 
      \includegraphics[width=0.24\textwidth]{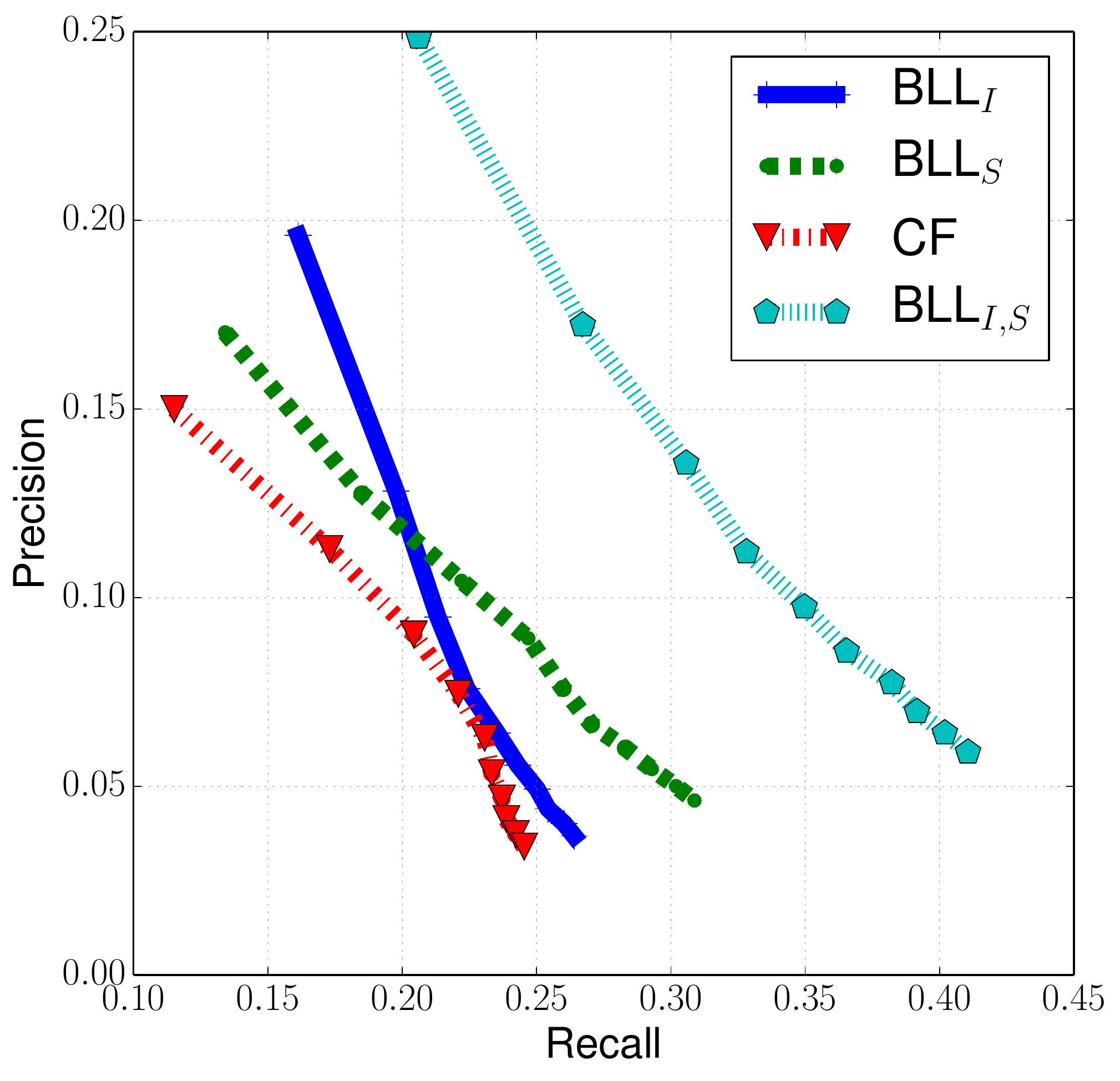} 
   }
     \subfloat[][\sone{}: Hashtag rec. w/o current tweet\\\rand{} dataset]{ 
      \includegraphics[width=0.24\textwidth]{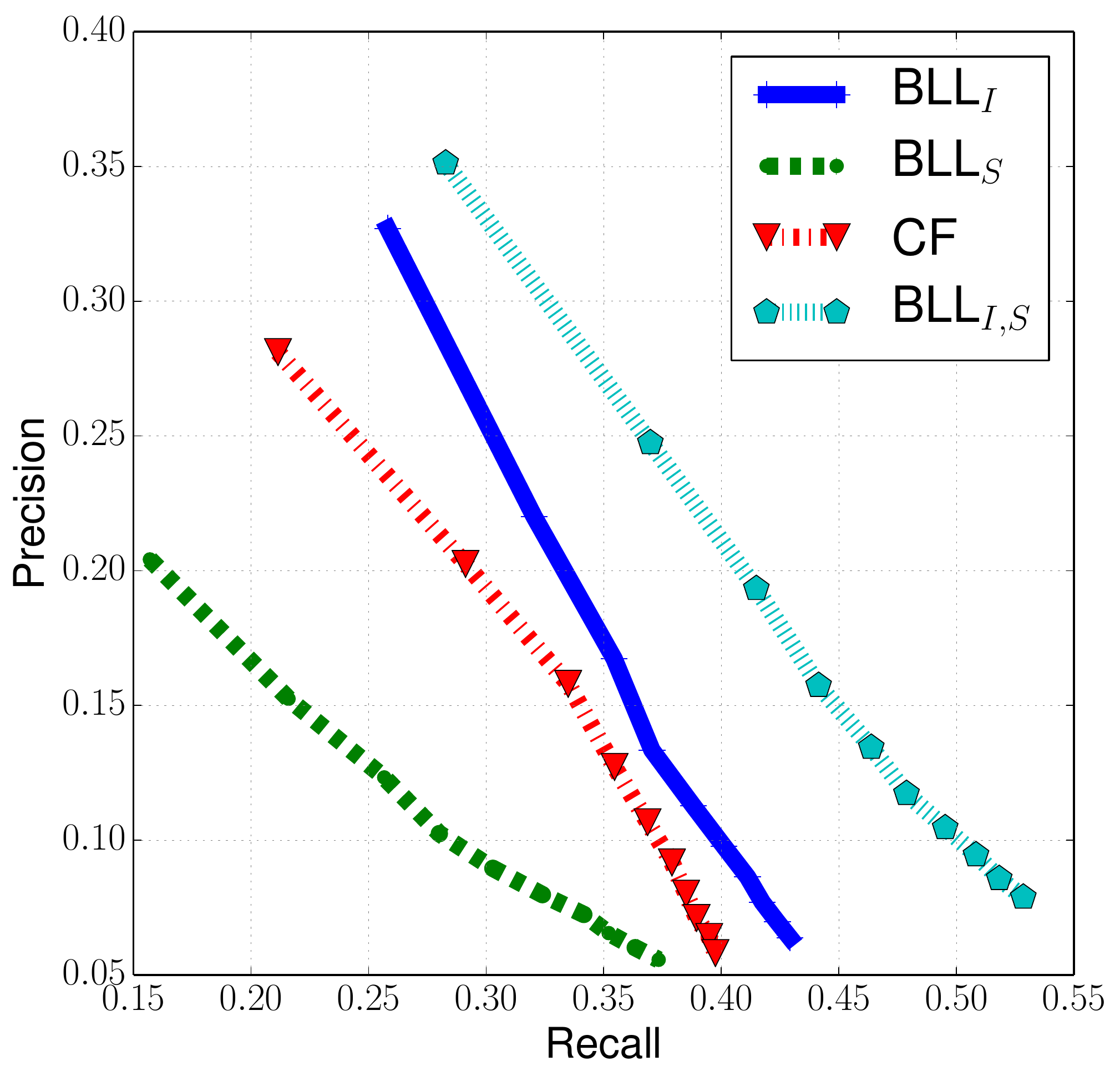} 
   }
     \subfloat[][\stwo{}: Hashtag rec. w/ current tweet\\\res{} dataset]{ 
      \includegraphics[width=0.24\textwidth]{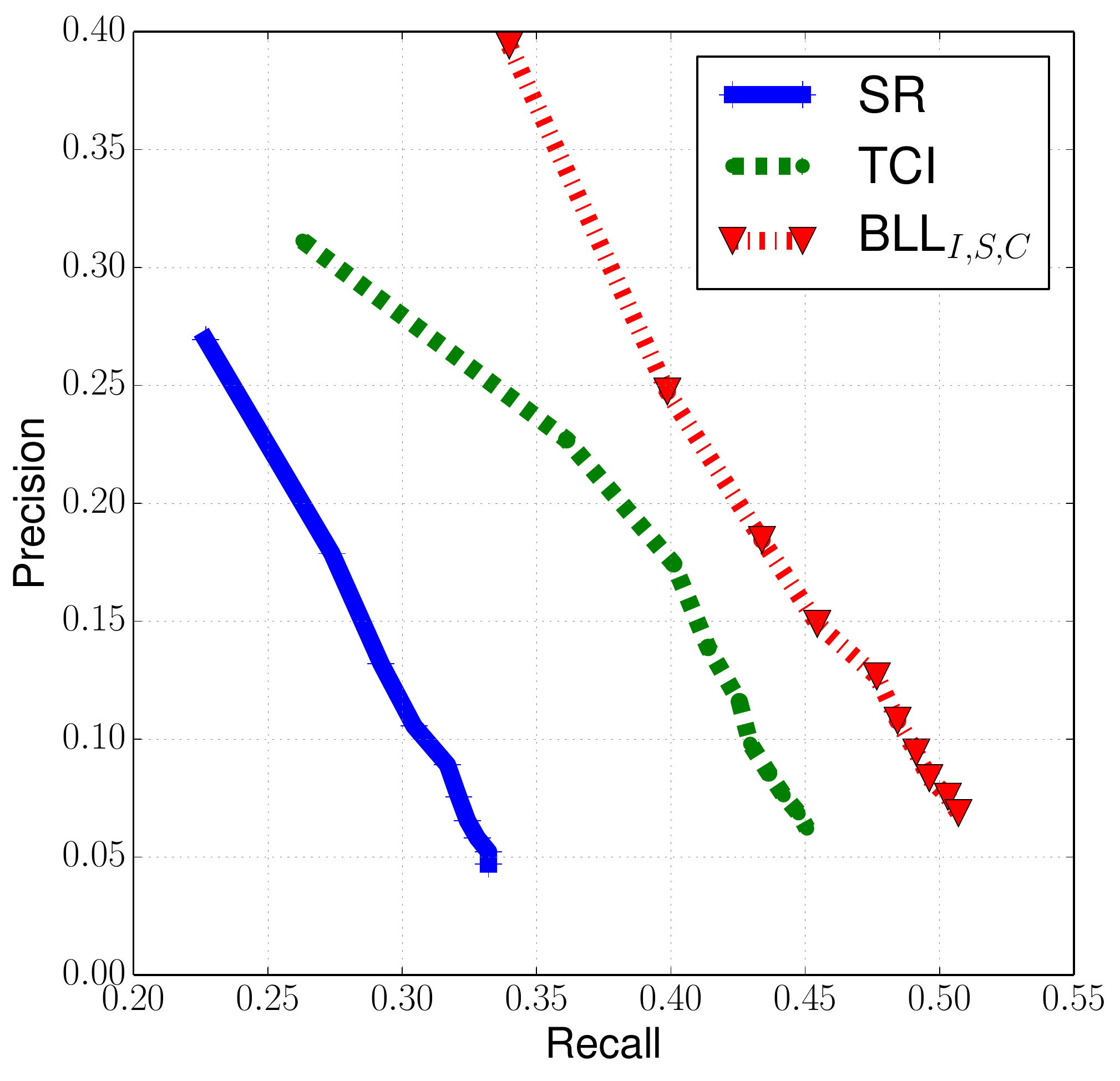} 
   }
     \subfloat[][\stwo{}: Hashtag rec. w/ current tweet\\\rand{} dataset]{ 
      \includegraphics[width=0.24\textwidth]{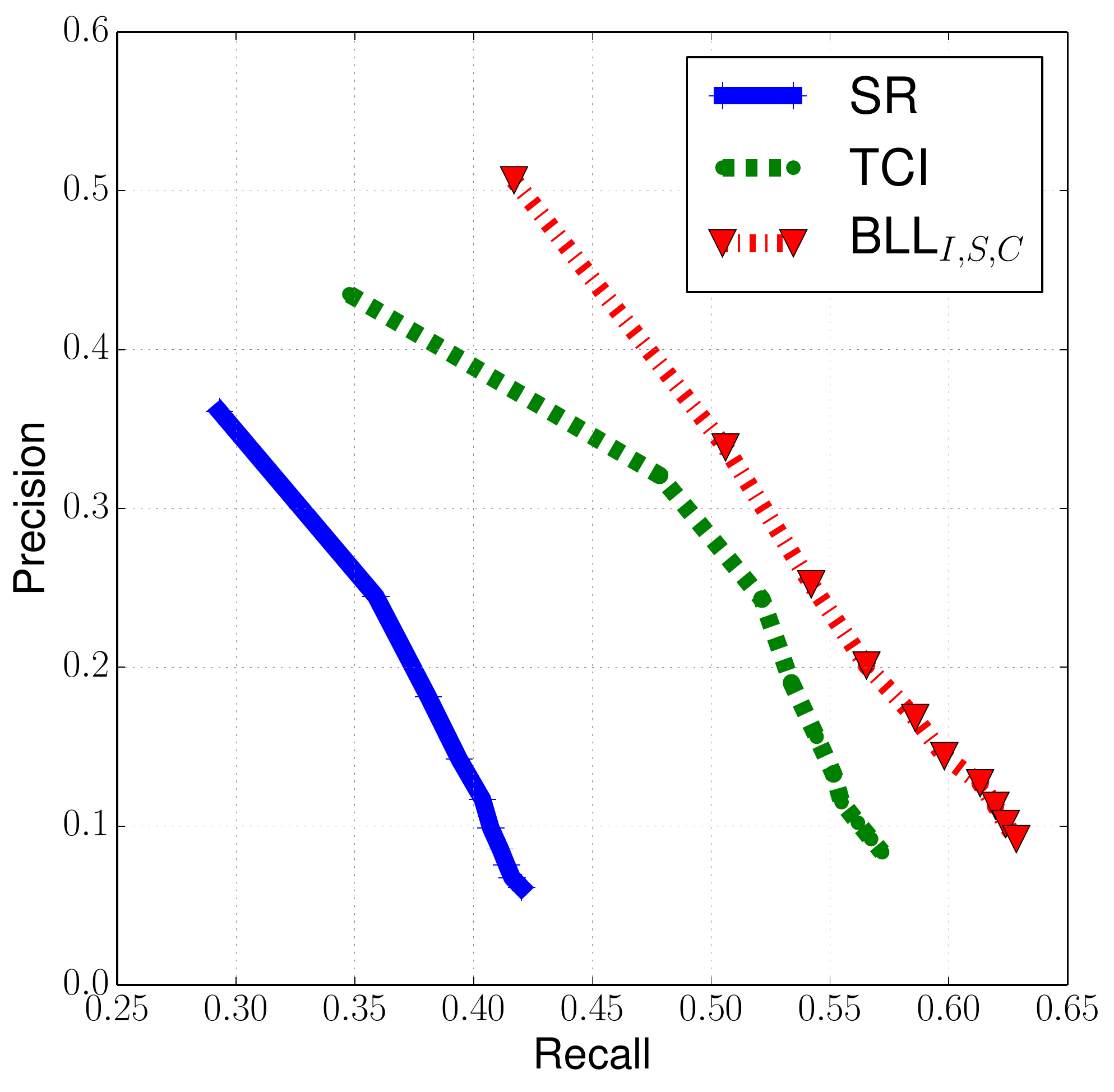} 
   }
   \caption{Precision / Recall plots of our evaluation scenarios for $k$ = 1 - 10 recommended hashtags.
\vspace{-2mm}}
     \label{fig:results}
\end{figure*}

\para{Conclusion and Reproducibility.} We find that temporal effects play an important role in hashtag reuse on Twitter. We propose our cognitive-inspired hashtag recommendation approaches \pred {} and \rec{} to account for such effects. We compare both algorithms to state-of-the-art hashtag recommendation algorithms and find that our cognitive-inspired approaches outperform these algorithms in terms of prediction accuracy and ranking. With our work, we aim to contribute to the rich line of research on improving the use of hashtags in social networks. We also hope to spark future work to utilize models from human memory theory to model and explain digital traces and user behavior online. 
For the sake of reproducibility, we implement and evaluate our approach by extending our open-source tag recommender benchmarking framework \textit{TagRec}. The source code and framework are freely accessible for scientific purposes on the Web\footnote{\url{https://github.com/learning-layers/TagRec}}.

\para{Keywords:} hashtag recommendation, 
ACT-R, 
temporal effects, 
hashtag reuse, 
user behavior modeling

\end{abstract}

\bibliography{references} 
\end{document}